\documentclass[twocolumn,showpacs,superscriptaddress,amsmath,amssymb,nofootinbib]{revtex4-1}
\usepackage{graphicx}
\usepackage{epsfig}
\usepackage{dcolumn}
\usepackage{multirow}
\usepackage{booktabs}
\usepackage{appendix}
\usepackage{bm}
\usepackage{overpic}
\usepackage{color}
\usepackage{subfigure}
\usepackage{threeparttable}
\usepackage[colorlinks,linkcolor=blue,anchorcolor=blue,citecolor=blue]{hyperref}

\def\Dp{D^{+}}
\def\Dm{D^{-}}

\def\pip{\pi^{+}}
\def\pim{\pi^{-}}
\def\piz{\pi^{0}}
\def\ks{K_{S}^{0}}
\def\ee{e^{+}e^{-}}
\def\dedx{\mathrm{d}E/\mathrm{d}x}
\def\e{\mathrm{e}}

\def\kp{K^+}

\def\BR{\mathcal{B}}

\def \gev  {\mbox{GeV}}
\def \gevc {\mbox{GeV/$c$}}
\def \gevcc{\mbox{GeV/$c^2$}}
\def \mev  {\mbox{MeV}}
\def \mevcc{\mbox{MeV/$c^2$}}

\def \ifb  {\mbox{fb$^{-1}$}}

\def \kpipi{K^{+}\pi^{-}\pi^{-}}
\def \kpipipiz{K^{+}\pi^{-}\pi^{-}\pi^{0}}
\def \kspipiz{\ks\pi^{-}\pi^{0}}
\def \kspi{\ks\pi^{-}}
\def \kspipipi{\ks\pi^{-}\pi^{-}\pi^{+}}
\def \kkpi{K^{+}K^{-}\pi^{-}}

\def \ss {\sqrt{s}}

\def \mbctag {M^{\rm{tag}}_{\rm{BC}}}

\def \dE {\Delta E}
\def \dEtag {\Delta E_{\rm{tag}}}
\def \dEsig {\Delta E_{\rm{sig}}}
\def \ebeam {E_{\rm{beam}}}
\def \mrec {M_{\rm{rec}}}
\def \mkppiz {M^{2}_{K^{+}\pi^{0}}}
\def \mkspiz {M^{2}_{K_{S}^{0}\pi^{0}}}

\def \kstarp {K^{*}(892)^{+}}
\def \kstarz {\bar{K}^*(892)^0}

\def \kppizswave {(K^{+}\pi^{0})_{\mathcal{S}\mathrm{-wave}}}
\def \kspizswave {(K_{S}^{0}\pi^{0})_{\mathcal{S}\mathrm{-wave}}}

\def \romanOne   {\uppercase\expandafter{\romannumeral1}}
\def \romanTwo   {\uppercase\expandafter{\romannumeral2}}
\def \romanThree {\uppercase\expandafter{\romannumeral3}}
\def \romanFour  {\uppercase\expandafter{\romannumeral4}}
\def \romanFive  {\uppercase\expandafter{\romannumeral5}}
\def \romanSix   {\uppercase\expandafter{\romannumeral6}}
\def \romanSeven {\uppercase\expandafter{\romannumeral7}}

\begin{document}
\title{\bf\boldmath Study of the decay $D^{+}\to K^{*}(892)^+K_S^0$  in $D^+\to K^+ K_S^0\pi^0$}
\author{\small
M.~Ablikim$^{1}$, M.~N.~Achasov$^{10,c}$, P.~Adlarson$^{67}$, S. ~Ahmed$^{15}$, M.~Albrecht$^{4}$, R.~Aliberti$^{28}$, A.~Amoroso$^{66A,66C}$, M.~R.~An$^{32}$, Q.~An$^{63,49}$, X.~H.~Bai$^{57}$
% ...
\\
\vspace{0.2cm}
(BESIII Collaboration)\\
\vspace{0.2cm} {\it
$^{1}$ Institute of High Energy Physics, Beijing 100049, People's Republic of China\\
% ...
$^{4}$ Bochum Ruhr-University, D-44780 Bochum, Germany\\
% ...
$^{10}$ G.I. Budker Institute of Nuclear Physics SB RAS (BINP), Novosibirsk 630090, Russia\\
% ...
$^{15}$ Helmholtz Institute Mainz, Staudinger Weg 18, D-55099 Mainz, Germany\\
% ...
$^{28}$ Johannes Gutenberg University of Mainz, Johann-Joachim-Becher-Weg 45, D-55099 Mainz, Germany\\
% ...
$^{32}$ Liaoning Normal University, Dalian 116029, People's Republic of China\\
% ...
$^{49}$ State Key Laboratory of Particle Detection and Electronics, Beijing 100049, Hefei 230026, People's Republic of China\\
% ...
$^{57}$ University of Jinan, Jinan 250022, People's Republic of China\\
% ...
$^{63}$ University of Science and Technology of China, Hefei 230026, People's Republic of China\\
% ...
$^{66}$ University of Turin and INFN, (A)University of Turin, I-10125, Turin, Italy; (B)University of Eastern Piedmont, I-15121, Alessandria, Italy; (C)INFN, I-10125, Turin, Italy\\
$^{67}$ Uppsala University, Box 516, SE-75120 Uppsala, Sweden\\
% ...
$^{c}$ Also at the Novosibirsk State University, Novosibirsk, 630090, Russia\\
% ...
}\vspace{0.4cm}}

\date{\today}
\begin{abstract}
Based on an $e^{+}e^{-}$ collision data sample corresponding to an integrated luminosity of 2.93 $\mathrm{fb}^{-1}$ collected with the BESIII detector at $\sqrt{s}=3.773\,\mathrm{GeV}$,
the first amplitude analysis of the singly Cabibbo-suppressed decay $D^+\to K^+ K_S^0\pi^0$ is performed.
From the amplitude analysis, the $K^{*}(892)^+K_S^0$ component is found to be dominant with a fraction of $(57.1\pm2.6\pm4.2)\%$, where the first uncertainty is statistical and the second systematic. In combination with the absolute branching fraction $\mathcal{B}(D^+\to K^+ K_S^0\pi^0)$ measured by BESIII, we obtain $\mathcal{B}(D^+\to K^{*}(892)^+K_S^0)=(8.69\pm0.40\pm0.64\pm0.51)\times10^{-3}$, where the third uncertainty is due to the branching fraction $\mathcal{B}(D^+\to K^+ K_S^0\pi^0)$. The precision of this result is significantly improved compared to the previous measurement. This result also differs from most of theoretical predictions by about $4\,\sigma$, which may help to improve the understanding of the dynamics behind.
\end{abstract}
\pacs{14.40.Lb, 13.20.Fc, 12.38.Qk}
\maketitle

	
\section{\boldmath Introduction}
The study of CP violation (CPV) in hadron decays is important for the understanding of the matter-antimatter asymmetry in the universe. In the charmed-meson sector, CPV effects in singly Cabibbo-suppressed (SCS) $D$-meson decays are usually much larger than those in Cabibbo-favored and doubly Cabibbo-suppressed decays. In 2019, the LHCb collaboration first observed a CPV effect in a combined analysis of $D^0\to \pip\pim$ and $D^0\to K^+K^-$ decays~\cite{Aaij:2019kcg}. However, theoretical predictions of this CPV effect suffer from large variations compared to those in the $K$ or $B$ meson systems, mainly due to the large uncertainty in describing the non-perturbative dynamics in QCD in the charm region~\cite{Saur:2020rgd, Ablikim:2019hff}.  

\begin{figure*}[tph]
\centering
\subfigure[\quad\quad\quad\quad\quad\quad]{
	\includegraphics[trim = 9mm 0mm 0mm 0mm, width=0.4\textwidth]{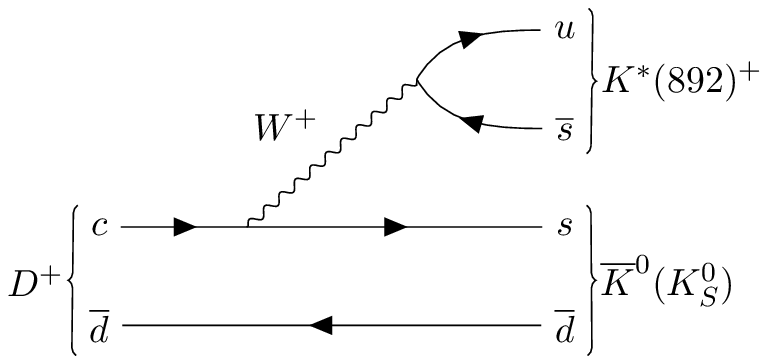}
}
\quad
\quad
\subfigure[\quad\quad\quad\quad\quad\quad]{
	\includegraphics[trim = 9mm 0mm 0mm 0mm, width=0.4\textwidth]{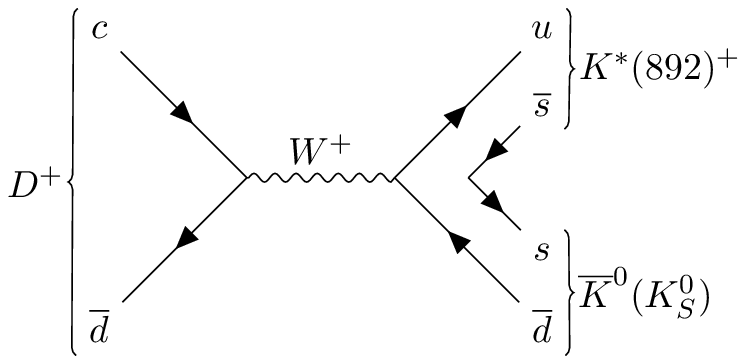}
}\\
\subfigure[\quad\quad\quad\quad\quad\quad]{
	\includegraphics[trim = 9mm 0mm 0mm 0mm, width=0.4\textwidth]{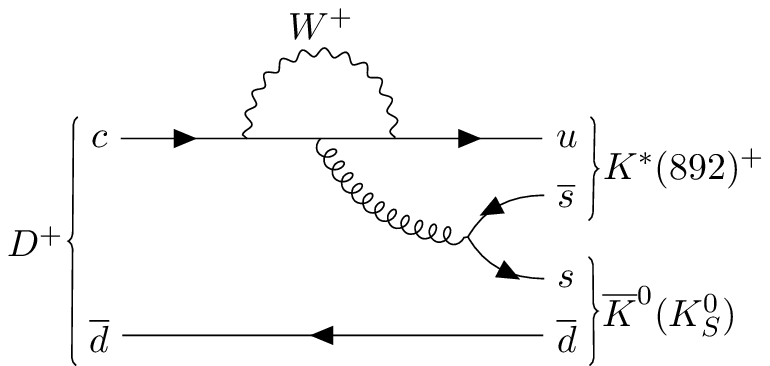}
}
\quad
\quad
\subfigure[\quad\quad\quad\quad\quad\quad]{
	\includegraphics[trim = 9mm 0mm 0mm 0mm, width=0.4\textwidth]{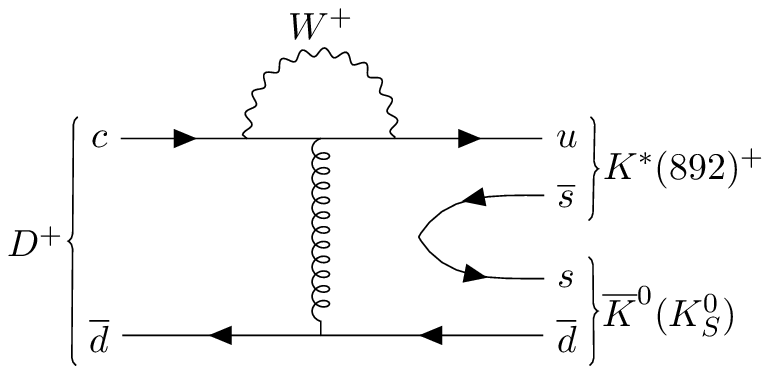}
}
\caption{Topological diagrams contributing to the decay $\Dp\to\kstarp\ks$ with (a) color-favored tree diagram, (b) $W$-annihilation diagram, (c) color-favored QCD penguin diagram and (d) QCD penguin exchange diagram.}
\label{fig:diagram}
\end{figure*}

In recent years, the branching fractions (BFs) and CPV in the two-body hadronic decays of $D\to PP$ and $D\to VP$ have been studied in different QCD-derived models~\cite{yufusheng2011,yufusheng2014,haiyang2016,haiyang2019}, where $P$ and $V$ denote pseudoscalar and vector mesons, respectively. Generally, these theoretical calculations are in good agreement with experimental results, except for those of the SCS decay $\Dp\to\kstarp\ks$, whose amplitude consists of color-favored tree diagrams, $W$-annihilation diagrams, and penguin diagrams~\cite{haiyang2019}. The topological diagrams can be found in Fig.~\ref{fig:diagram}.
The measured and predicted values of $\BR(\Dp\to\kstarp\ks)$ are listed in Table~\ref{tab:summation}. The E687 collaboration reported a BF ratio of $\frac{\BR(\Dp\to\kstarp(\to\kp\piz)\ks)}{\BR(\Dp\to\ks\pip)}=1.1\pm0.3\pm0.4$~\cite{quote1}.
This results in $\BR(\Dp\to\kstarp\ks) = (17\pm8)\times10^{-3}$ when combined with the world average of $\BR(\Dp\to\ks\pip)$~\cite{pdg2020}.
Although the predicted values are consistent with the experimental results, the experimental precision needs to be improved.
A precise measurement of $\BR(\Dp\to\kstarp\ks)$ will provide a more stringent test of the theoretical models and help to deepen our understanding of the dynamics of charmed meson decays. Especially, this will enhance the predictive power on the CPV in charmed meson decays.

In this paper, the first amplitude analysis of the SCS decay $\Dp\to\kp\ks\piz$ is reported, based on a sample of $\Dp\Dm$ pairs from $\ee$ collisions at a center-of-mass energy of $\ss=3.773\,\gev$ corresponding to an integrated luminosity of 2.93 $\ifb$~\cite{lumi,Ablikim:2015orh} collected with the BESIII detector~\cite{detector} at the Beijing Electron Positron Collider (BEPCII)~\cite{BEPCII}. At the energy $\ss=3.773\,\gev$, the pairs of $\Dp\Dm$ are produced near threshold without any accompanying hadron~\cite{Li:2021iwf}.
Previously, the BESIII collaboration has measured the BF of $\Dp\to\kp\ks\piz$ to be $(5.07\pm0.19\pm0.23)\times10^{-3}$~\cite{quote3}.
In combination with the amplitude analysis results presented in this paper, the BF of $\Dp\to\kstarp\ks$ can be determined with much improved precision. 
Charge conjugation is implied throughout the text.

\begin{table}[!htp]
\caption{Predicted BFs of the decay $\Dp\to\kstarp\ks$ from the pole model~\cite{yufusheng2011}, the factorization-assisted topological-amplitude (FAT) approach with $\rho$-$\omega$ mixing~\cite{yufusheng2014}, the topological diagram approach with only tree level amplitude (denoted as TDA[tree])~\cite{haiyang2016}, and including QCD-penguin amplitudes (denoted as TDA[QCD-penguin])~\cite{haiyang2019}. For comparison, the previous experimental result~\cite{quote1,pdg2020} is also listed.}
\label{tab:summation}
\begin{center}
\begin{tabular}{c|c}
\hline\hline
Model & $\BR(\Dp\to\kstarp\ks)$($\times10^{-3}$) \\
\hline
Pole & $6.2\pm1.2$\\
FAT[mix] & $5.5$\\
TDA[tree] & $5.02 \pm 1.31$\\
TDA[QCD-penguin] & $4.90\pm0.21$\\
\hline
PDG & $17\pm8$\\
\hline\hline
\end{tabular}
\end{center}
\end{table}

\section{\boldmath BESIII Experiment and Monte Carlo Simulation}
The BESIII detector records symmetric $\ee$ collisions provided by the BEPCII storage ring, which operates with a peak luminosity of $1\times10^{33}$~cm$^{-2}$s$^{-1}$ in the center-of-mass energy range from 2.0 to $4.9\,\gev$. BESIII has collected large data samples in this energy region~\cite{Ablikim:2019hff}. The cylindrical core of the BESIII detector covers 93\% of the full solid angle and consists of a helium-based multilayer drift chamber~(MDC), a plastic scintillator time-of-flight system (TOF), and a CsI(Tl) electromagnetic calorimeter (EMC), which are all enclosed in a superconducting solenoidal magnet providing a 1.0 T magnetic field. The solenoid is supported by an octagonal flux-return yoke with resistive plate counter muon identification modules interleaved with steel. The charged-particle momentum resolution at $1\,\gevc$ is $0.5\%$, and the ionization energy loss $\dedx$ resolution is $6\%$ for electrons from Bhabha scattering. The EMC measures photon energies with a resolution of $2.5\%$ ($5\%$) at $1\,\gev$ in the barrel (end cap) region. The time resolution in the TOF barrel region is 68\,ps, while that in the end cap region is 110\,ps. More detailed descriptions can be found in Refs.~\cite{detector,BEPCII}.

Simulated data samples produced with a Geant4-based~\cite{geant4} Monte Carlo (MC) package, which includes the geometric description of the BESIII detector~\cite{GDMLMethod,BesGDML} and the detector response, are used to estimate background contributions and obtain the reconstruction efficiency. The simulation models the beam energy spread and initial state radiation (ISR) in the $\ee$ annihilations with the generator {\sc kkmc}~\cite{kkmc}.
The `inclusive MC sample' includes the production of $D\bar{D}$ pairs (including quantum coherence for the neutral $D$ channels), non-$D\bar{D}$ decays of the $\psi(3770)$, ISR production of the $J/\psi$ and $\psi(3686)$ states, and continuum processes which are incorporated in {\sc kkmc}~\cite{kkmc}. Known decay modes are modeled with {\sc evtgen}~\cite{evtgen} using the BFs published by the Particle Data Group (PDG)~\cite{pdg2020}, and the remaining unknown charmonium decays are modeled with {\sc lundcharm}~\cite{lundcharm}. The final state radiation from charged final state particles is incorporated using {\sc photos}~\cite{photos}. The inclusive MC sample is used to study background contributions and to estimate signal purity. In this work, two sets of signal MC samples are used. One sample is generated with a uniform distribution in phase space (PHSP) for the decay $\Dp\to\kp\ks\piz$, called the `PHSP MC sample', which is used to extract the detection efficiency maps along the Dalitz plot coordinates. The other sample is generated based on the fitted amplitudes from the amplitude analysis, called the `DIY MC sample'. It is used to evaluate the fit quality and estimate the systematic uncertainty. The recoiling $\Dm$ in these two sets of MC samples is forced to decay into six tag modes, discussed in Sec.~\ref{sec:tagsel}.

\section{\boldmath Event Selection}
\label{sec:selection}
Taking advantage of the threshold production of the $\Dp\Dm$ sample, this analysis uses a double-tag method, which is illustrated in the following.

\subsection{\boldmath Tagged candidate selection}
\label{sec:tagsel}
The six tag modes used to tag $\Dm$ candidates are $\kpipi$, $\kpipipiz$, $\kspi$, $\kspipiz$, $\kspipipi$ and $\kkpi$, with subsequent $\piz\to\gamma\gamma$ and $\ks\to\pip\pim$ decays. The sum of their BFs is about 27.7\% ~\cite{pdg2020}. The tagged candidates are reconstructed from all possible combinations of final state particles according to the following selection criteria.

Charged particle tracks are reconstructed using the information of the MDC, and are required to have a polar angle $\theta$ with respect to the $z$-axis, defined as the symmetry axis of the MDC, satisfying $|\!\cos\theta|<0.93$ and to have a distance of closest approach to the interaction point (IP) smaller than 10 cm along the $z$-axis ($V_z$) and smaller than 1 cm in the perpendicular plane ($V_r$). Those tracks used in reconstructing $\ks\to\pip\pim$ decays are exempt from these selection criteria. Particle identification (PID) for charged particle tracks is implemented by using combined information from the flight time measured in the TOF and the $\dedx$ measured in the MDC to form a PID probability $\mathcal{L}(h)$ for each hadron $(h)$ hypothesis with $h=\pi,K$. Charged tracks are identified as pions when they satisfy $\mathcal{L}(\pi)>\mathcal{L}(K)$, and as kaons otherwise.

Photon candidates from $\piz$ decays are reconstructed from the electromagnetic showers detected in the EMC crystals. The deposited energy is required to be larger than $25\,\mev$ in the barrel region with $|\!\cos\theta|<0.80$ and larger than $50\,\mev$ in the end cap region with $0.86<|\!\cos\theta|<0.92$. To further suppress fake photon candidates due to electronic noise or beam background, the measured EMC time is required to be within 700 ns from the event start time. To reconstruct $\piz$ candidates, the invariant mass of a photon pair is required to satisfy $0.115<M_{\gamma\gamma}<0.150\,\gevcc$. To further improve the momentum resolution, the invariant mass of the photon pair is constrained to the nominal $\piz$ mass~\cite{pdg2020} by applying a one-constraint kinematic fit. The updated momentum of the $\piz$ is used in the further analysis.

$\ks$ candidates are reconstructed through the decay $\ks\to\pip\pim$ by combining all pairs of oppositely charged tracks, without applying the PID requirement. These tracks need to satisfy $|\!\cos\theta|<0.93$ and $V_z<20\,\mathrm{cm}$ while no $V_r$ requirement is applied. A vertex fit is applied to pairs of charged tracks constraining them to originate from a common decay vertex, and the $\chi^2$ of this vertex fit is required to be less than 100. The invariant mass of the $\pip\pim$ pair needs to satisfy $0.487<M_{\pip\pim}<0.511\,\gevcc$.  Here, $M_{\pip\pim}$ is calculated with the pions constrained to originate at the decay vertex.

To identify the tagged candidates, two kinematic variables, the beam-constrained mass $\mbctag$, and the energy difference $\dEtag$, are defined as
\begin{eqnarray}
\mbctag \equiv \sqrt{{\ebeam}^2/c^4-\left|\vec{p}_{\rm{tag}}\right|^2/c^2},
\label{eq:mbctag}
\end{eqnarray}
and
\begin{eqnarray}
\dEtag \equiv E_{\rm{tag}}-\ebeam ,
\label{eq:dEtag}
\end{eqnarray}
where $\vec{p}_{\rm{tag}}$ and $E_{\rm{tag}}$ are the three-momentum and energy of the tagged $\Dm$ candidate in the rest frame of the initial $\ee$ collision system, and $\ebeam$ is the beam energy. For a correctly reconstructed tagged candidate, $\mbctag$ and $\dEtag$ are expected to be consistent with the nominal $\Dm$ mass~\cite{pdg2020} and zero, respectively.
In this work, the same $\dEtag$ requirements as those in a previous BESIII analysis~\cite{zhangyu} are used, which are listed in Table~\ref{tab:tagMode}. In each event, only the combination with the smallest $|\dEtag|$ is kept for each tag mode. The tagged candidates are required to be within the region $1.863<\mbctag<1.879\,\gevcc$.

\begin{table}[htbp]
\begin{center}
\caption{$\dEtag$ requirements for different $\Dm$ tag modes.}
\begin{tabular}{l c}
      \hline \hline
      $\Dm$ decay & $\dEtag(\gev)$\\
      \hline
      $\kpipi$ & $(-0.022,0.021)$\\
      $\kpipipiz$ & $(-0.060,0.034)$\\
      $\kspi$ & $(-0.019,0.021)$\\
      $\kspipiz$ & $(-0.071,0.041)$\\
      $\kspipipi$ & $(-0.025,0.023)$\\
      $\kkpi$ & $(-0.019,0.018)$\\
       \hline\hline
\end{tabular}
\label{tab:tagMode}
\end{center}
\end{table}

\subsection{\boldmath Signal candidate selection}
\label{sec:sigsel}
Signal candidates for $\Dp\to\kp\ks\piz$ are formed using the remaining tracks recoiling against the tagged $\Dm$. Besides the selection requirements for charged and neutral tracks described in Sec.~\ref{sec:tagsel}, some additional criteria are applied to improve the signal-to-background ratio.
For the $\ks$ candidates, an additional secondary vertex fit is applied where the momentum of the reconstructed $\ks$ candidate is constrained to be aligned with the direction from the IP to the $\ks$ decay vertex, and the resulting flight length $L$ is required to be larger than twice its uncertainty $\sigma_L$. The $\chi^2$ of the secondary vertex fit is required to be less than 500.
For the $\piz$ candidates, the $\chi^2$ of the kinematic fit is required to be less than 20.

To further identify the signal $\Dp$ candidates, the energy difference, $\dEsig$, is defined as
\begin{eqnarray}
\dEsig \equiv E_{\rm{sig}}-\ebeam,
\label{eq:dEtag}
\end{eqnarray}
where $E_{\rm{sig}}$ is the energy of the signal $\Dp$ candidate in the rest frame of the initial $\ee$ collision system. In each event, only the combination with the least $|\dEsig|$ is kept as a signal candidate. The $\dEsig$ distributions of data and inclusive MC sample are shown in Fig.~\ref{fig:distribution}(a). We require $-0.03<\dEsig<0.02\,\gev$ for signal candidates to be kept. 
To further improve the momentum resolution of the signal final state $\kp\ks\piz$, an additional kinematic fit is applied constraining the invariant mass of the signal final state to the nominal $\Dp$  mass~\cite{pdg2020}, and the total four-momentum of all reconstructed particles to the initial $\ee$ collision four-momentum. The updated four momenta are used for further analysis.

\begin{figure}[tph]
\centering
\includegraphics[trim = 9mm 0mm 0mm 0mm, width=0.4\textwidth]{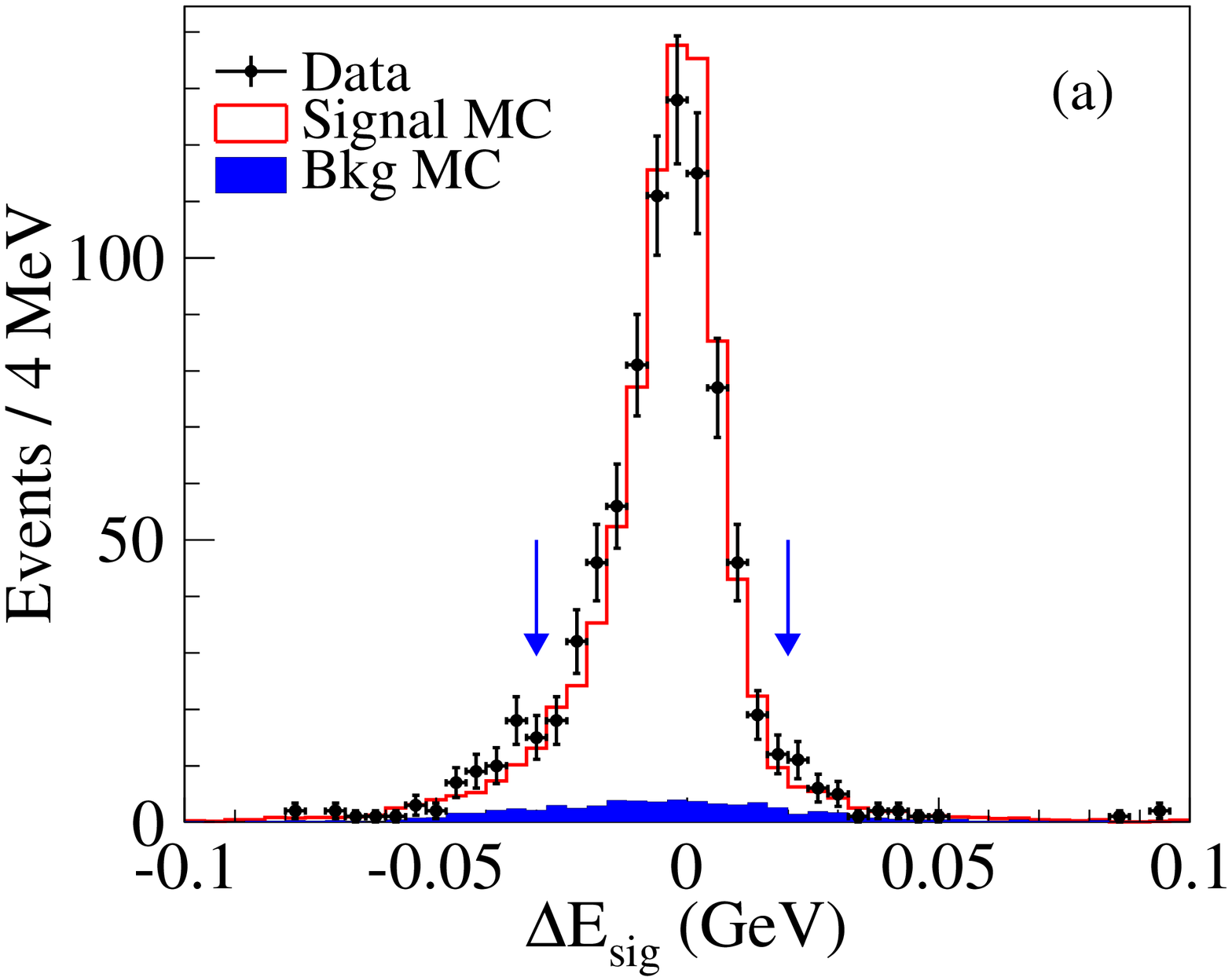}\\
\includegraphics[trim = 9mm 0mm 0mm 0mm, width=0.4\textwidth]{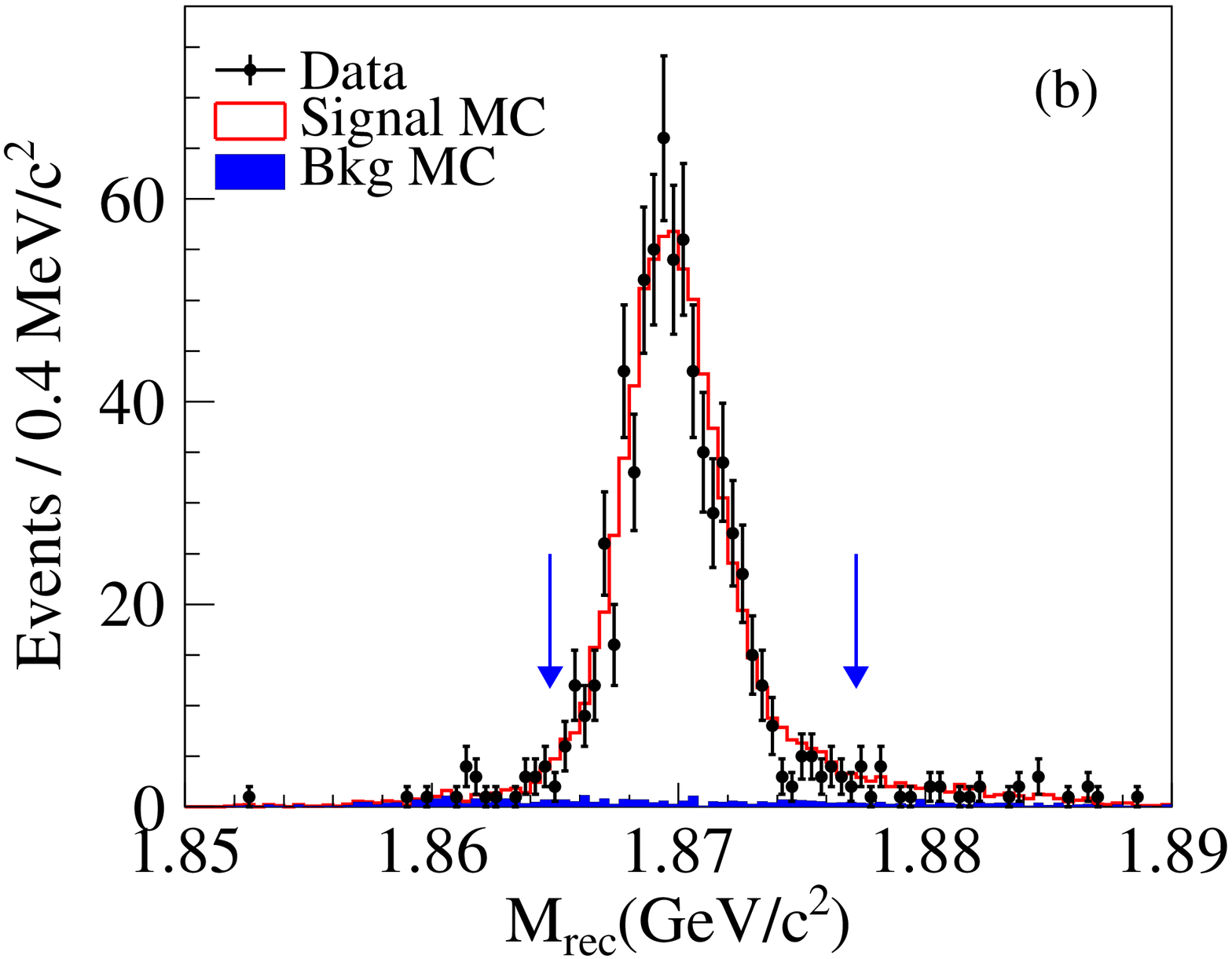}
\caption{Distributions of $\dE_{\rm{sig}}$ (a) and $\mrec$ (b) in data and the scaled inclusive MC sample. The points with uncertainties denote data and the unshaded (shaded) histogram denotes the signal (background) events from the scaled inclusive MC sample. The arrows indicate the $\dE_{\rm{sig}}$ and $\mrec$ requirements.}
\label{fig:distribution}
\end{figure}

The recoil mass $\mrec$ is defined as
\begin{eqnarray}
\mrec \equiv \sqrt{(\bm{p}_{\ee}-\bm{p}_{\Dp})^2}/c,
\label{eq:m_recoil}
\end{eqnarray}
where $\bm{p}_{\ee}$ is the $\ee$ collision initial four-momentum and $\bm{p}_{\Dp}$ is the four-momentum of the $\Dp$ signal candidate. The distribution of $\mrec$ in data and in the inclusive MC sample is shown in Fig.~\ref{fig:distribution}(b). 
The candidate events within $1.865<\mrec<1.877\,\gevcc$ are selected and the signal purity is determined to be $(97.4\pm 0.2)\%$ according to the inclusive MC sample. After imposing all above selection criteria, the number of the signal events in data is measured to be 692.

\section{Amplitude analysis}
Only spin-zero particles are involved in the signal process $\Dp\to\kp\ks\piz$, 
thus only two degrees of freedom are needed to describe the full kinematics. In the amplitude analysis, we choose the two Dalitz plot variables $\mkppiz$ and $\mkspiz$. 

\subsection{Isobar model}
The full amplitude of the decay process is described by the Isobar model~\cite{radius_ori}, which is given by
\begin{eqnarray}
\mathcal{M}(\Dp\to\kp\ks\piz)=\sum_i{c_i\cdot\mathcal{A}_i},
\label{eq:full_amp}
\end{eqnarray}
where the term $c_i=a_i\e^{i\phi_i}$ consists of the magnitude $a_i$ and the phase $\phi_i$ for the specific intermediate process $i$. The amplitude $\mathcal{A}_i$ denotes the decay amplitude of the process $i$, which is modeled in a quasi-two-body decay $D\to Cr$, $r\to AB$. Here, $r$ is the possible resonance decaying into $AB$, and $A$, $B$ and $C$ each denote one of the final state particles $\kp\ks\piz$. $\mathcal{A}_i$ is formulated as
\begin{eqnarray}
\mathcal{A}_i=F_D\times T_i\times F_r\times W_i,
\label{eq:twobody_amp}
\end{eqnarray}
where $W_i$ is the spin factor, $F_r$ and $F_D$ are the Blatt-Weisskopf barrier factors~\cite{barrier_factor}, and $T_i$ is the dynamical function describing the intermediate resonance, as illustrated below.

\subsubsection{Spin factor}
The spin factor $W_i$ for the process $D\to Cr$, $r\to AB$ is constructed based on the Zemach formalism~\cite{Zemach}. The amplitudes for resonances with angular momenta larger than two are not considered due to the limited phase-space. 
The spin factor is expressed as
\begin{eqnarray}
\small
\begin{aligned}
\mathrm{Spin\mbox{-}0}: W= &\;1,\\
\mathrm{Spin\mbox{-}1}: W= &\;M^2_{BC}-M^2_{AC}+\frac{(M^2_D-M^2_C)(M^2_A-M^2_B)}{M^2_{AB}},\\
\mathrm{Spin\mbox{-}2}: W= &\;a_1-\frac{1}{3}a_2a_3,
\end{aligned}
\end{eqnarray}
where $a_1$, $a_2$, and $a_3$ are given by
\begin{eqnarray}
\small
\begin{aligned}
a_1=&\;\left[M^2_{BC}-M^2_{AC}+\frac{(M^2_D-M^2_C)(M^2_A-M^2_B)}{M^2_{AB}}\right]^2,\\
a_2=&\;M^2_{AB}-2M^2_D-2M^2_C+\frac{\left(M^2_D-M^2_C\right)^2}{M^2_{AB}},\\
a_3=&\;M^2_{AB}-2M^2_A-2M^2_B+\frac{\left(M^2_A-M^2_B\right)^2}{M^2_{AB}}.
\end{aligned}
\end{eqnarray}
Here, $M_{AB}$, $M_{AC}$, and $M_{BC}$ denote the invariant masses of the particle combinations $AB$, $AC$, and $BC$, respectively, and $M_{A}$, $M_{B}$, $M_{C}$, and $M_{D}$ denote the nominal masses of $A$, $B$, $C$, and $D$, respectively.

\subsubsection{Blatt-Weisskopf barrier factors}
The Blatt-Weisskopf barrier factors $F_D$ and $F_r$ attempt to model the underlying quark structure of the parent particle in the decay $D\to rC$ and the subsequent decay $r\to AB$, respectively. The expressions for the barrier factors, which are shown in Table~\ref{tab:barrier_factor}, are taken from Ref.~\cite{barrier_factor}.
In these expressions, $p$ is the decay momentum of the particle $A$ ($C$) in the rest frame of the mother particle $r$ ($D$), while $q$ is the decay momentum of the particle $A$ ($C$) in the rest frame of the mother particle $r$ ($D$) when the resonance is fixed at the corresponding nominal mass. The radii of $\Dp$ and the intermediate resonance $r$ are chosen as $R_D=5\,\gev^{-1}$ and $R_r=1.5\,\gev^{-1}$, respectively~\cite{radius_ori,radius}.

\begin{table}[htbp]
\begin{center}
\caption{Expressions for Blatt-Weisskopf barrier factors~\cite{barrier_factor} under different angular momenta $L$. }
\begin{tabular}{c | c  c}
      \hline \hline
      ~~~~~~$L$~~~~~~ & $F$\\
      \hline
      0 &  $1$\\
      1 &  $\sqrt{\frac{1+(R\cdot p)^2}{1+(R\cdot q)^2}}$\\
      2 &  $\sqrt{\frac{9+3(R\cdot p)^2+(R\cdot p)^4}{9+3(R\cdot q)^2+(R\cdot q)^4}}$\\
      \hline\hline
\end{tabular}
\label{tab:barrier_factor}
\end{center}
\end{table}

\subsubsection{Dynamical function}
The dynamical function describes the line-shape of the intermediate resonance. For the specific process $r\to AB$, the dynamical function is chosen as a relativistic Breit-Wigner function for most of the resonances and is written as
\begin{eqnarray}
T(M_{AB})=\frac{1}{M^2_r-M^2_{AB}-iM_r\Gamma(M_{AB})},
\label{eq:RBW}
\end{eqnarray}
where $M_r$ is the nominal resonance mass, $M_{AB}$ is the invariant mass of $AB$ and $\Gamma(M_{AB})$ is the mass-dependent width defined as
\begin{eqnarray}
\Gamma(M_{AB})=\Gamma_r\left(\frac{p_{AB}}{p_r}\right)^{2J+1}\left(\frac{M_r}{M_{AB}}\right)F_r^2,
\label{eq:width}
\end{eqnarray}
where $\Gamma_r$ is the nominal resonance width, $J$ is the spin of the resonance and $p_{AB}$ and $p_r$ are the breakup momenta at $M_{AB}$ and $M_{r}$, respectively.

For the dynamical function of the $K\pi$ $\mathcal{S}$-wave, we choose the LASS parametrization~\cite{LASS}.
It includes both the $K^{*}_{0}(1430)$ resonances and a non-resonant part. We denote the full LASS parametrization as $\kppizswave$ or $\kspizswave$ in the following. The LASS parametrization can be expressed as
\begin{eqnarray}
\begin{aligned}
T(M_{K\pi})=&\frac{M_{K\pi}}{p_{K\pi}}\cdot[F_{K\pi}^{\mathrm{NR}}\cdot\sin(\delta_{K\pi}^{\mathrm{NR}}+\phi_{K\pi}^{\mathrm{NR}})\cdot\e^{i(\delta_{K\pi}^{\mathrm{NR}}+\phi_{K\pi}^{\mathrm{NR}})} \\
&+F_{K\pi}^{K_0^*}\cdot\sin\delta_{K\pi}^{K_0^*}\cdot\e^{i(\delta_{K\pi}^{K_0^*}+\phi_{K\pi}^{K_0^*})}\cdot\e^{2i(\delta_{K\pi}^{\mathrm{NR}}+\phi_{K\pi}^{\mathrm{NR}})}],
\end{aligned}
\label{eq:LASS}
\end{eqnarray}
where $F_{K\pi}^{\mathrm{NR}}$ ($\phi_{K\pi}^{\mathrm{NR}}$) and $F_{K\pi}^{K_0^*}$ ($\phi_{K\pi}^{K_0^*}$) are the magnitudes (phases) 
for the non-resonant and  $K_0^*(1430)$ components, respectively, and $p_{K\pi}$ is the breakup momentum of the $K\pi$ system. The phase-shifts $\delta_{K\pi}^{\mathrm{NR}}$ and $\delta_{K\pi}^{K_0^*}$ are defined as
\begin{eqnarray}
\delta_{K\pi}^{\mathrm{NR}}=\cot^{-1}\left(\frac{1}{a_{\mathrm{scat}}p_{K\pi}}+\frac{r_{\mathrm{eff}}p_{K\pi}}{2}\right),
\label{eq:deltaB}
\end{eqnarray}
and
\begin{eqnarray}
\delta_{K\pi}^{K_0^*}=\tan^{-1}\left[\frac{M_{K_0^*}\Gamma(M_{K\pi})}{M_{K_0^*}^2-M^2_{K\pi}} \right],
\label{eq:deltaB}
\end{eqnarray}
where $a_{\mathrm{scat}}$ is the scattering length, $r_{\mathrm{eff}}$ is the effective interaction range, and $M_{K_0^*}$ is the nominal mass of the $K^{*}_0(1430)$. The LASS parametrization corresponds to a $K$-matrix approach~\cite{Kmatrix} describing a rapid phase shift coming from the resonant term and a slowly rising shift governed by the non-resonant term, with relative strengths $F_{K\pi}^{K_0^*}$ and $F_{K\pi}^{\mathrm{NR}}$.
In the nominal fit, the LASS parameters are fixed according to the values measured by the BaBar and Belle collaborations~\cite{LASSpar}, which are listed in Table~\ref{tab:LASS}.

\begin{table}[htbp]
\renewcommand{\arraystretch}{1.14}
\begin{center}
\caption{Parameters of the $K\pi$ $\mathcal{S}$-wave component measured by BaBar and Belle~\cite{LASSpar}.}
\begin{tabular}{lcc}
      \hline \hline
      Parameter &  Value & Unit\\
      \hline
      $M_{K^*_0}$ & $1.441\pm0.002$ & $\gevcc$ \\
      $\Gamma_{K^*_0}$ & $0.193\pm0.004$ & $\gev$ \\
      $F_{K\pi}^{\mathrm{NR}}$ & $0.96\pm 0.07$ & ---\\
      $\phi_{K\pi}^{\mathrm{NR}}$ & $0.1\pm0.3$ & deg.\\
      $F_{K\pi}^{K_0^*}$ & 1 (fixed) & ---\\
      $\phi_{K\pi}^{K_0^*}$ & $-109.7\pm2.6$ & deg. \\
      $a_{\mathrm{scat}}$ & $0.113\pm0.006$ & $(\mathrm{GeV}/c)^{-1}$\\
      $r_{\mathrm{eff}}$ & $-33.8\pm1.8$ & $(\mathrm{GeV}/c)^{-1}$\\
      \hline\hline
\end{tabular}
\label{tab:LASS}
\end{center}
\end{table}

\subsection{Likelihood function}
In the amplitude analysis, a maximum likelihood fit is performed by minimizing the negative log-likelihood (NLL), which is constructed on the Dalitz plot plane as
\begin{footnotesize}
\begin{eqnarray}
-\ln{\mathcal{L}}=-\sum_{\mathrm{events}}\ln\left[
\eta\left(x,y\right)\cdot
\frac{\sum_{i,j}{c_ic_j^*\mathcal{A}_i\left(x,y\right)\mathcal{A}^*_j\left(x,y\right)}}{\sum_{i,j}{c_ic_j^*I_{ij}}}
\right],
\label{eq:likelihood}
\end{eqnarray}
\end{footnotesize}

\noindent where $(x,y)$ denote the Dalitz plot coordinates $(\mkppiz,\mkspiz)$, $\eta(x,y)$ is the efficiency function based on the smoothed histogram (following the method in Ref.~\cite{histpdf}) from the PHSP MC sample, where the Dalitz plot and the corresponding projections of the PHSP MC are illustrated in Fig.~\ref{fig:eff}, $\mathcal{A}_i(x,y)$ is the decay amplitude of the $i$-th component in Eq.~\eqref{eq:twobody_amp}, $c_i$ is the free complex coefficient of the $i$-th component, and $I_{ij}$ is the normalization integral, which is defined as
\begin{eqnarray}
I_{ij}=\int\mathcal{A}_i\left(x,y\right)\mathcal{A}^*_j\left(x,y\right) \eta\left(x,y\right)\mathrm{d}x\mathrm{d}y.
\label{eq:norm}
\end{eqnarray}
Here, the integral is calculated numerically by dividing the Dalitz plot plane into a grid of $3500\times3500$ square cells. No background contribution is included in the NLL in the nominal fit, exploiting the high signal purity.

\begin{figure*}[htp]\centering
\includegraphics[width=0.33\textwidth]{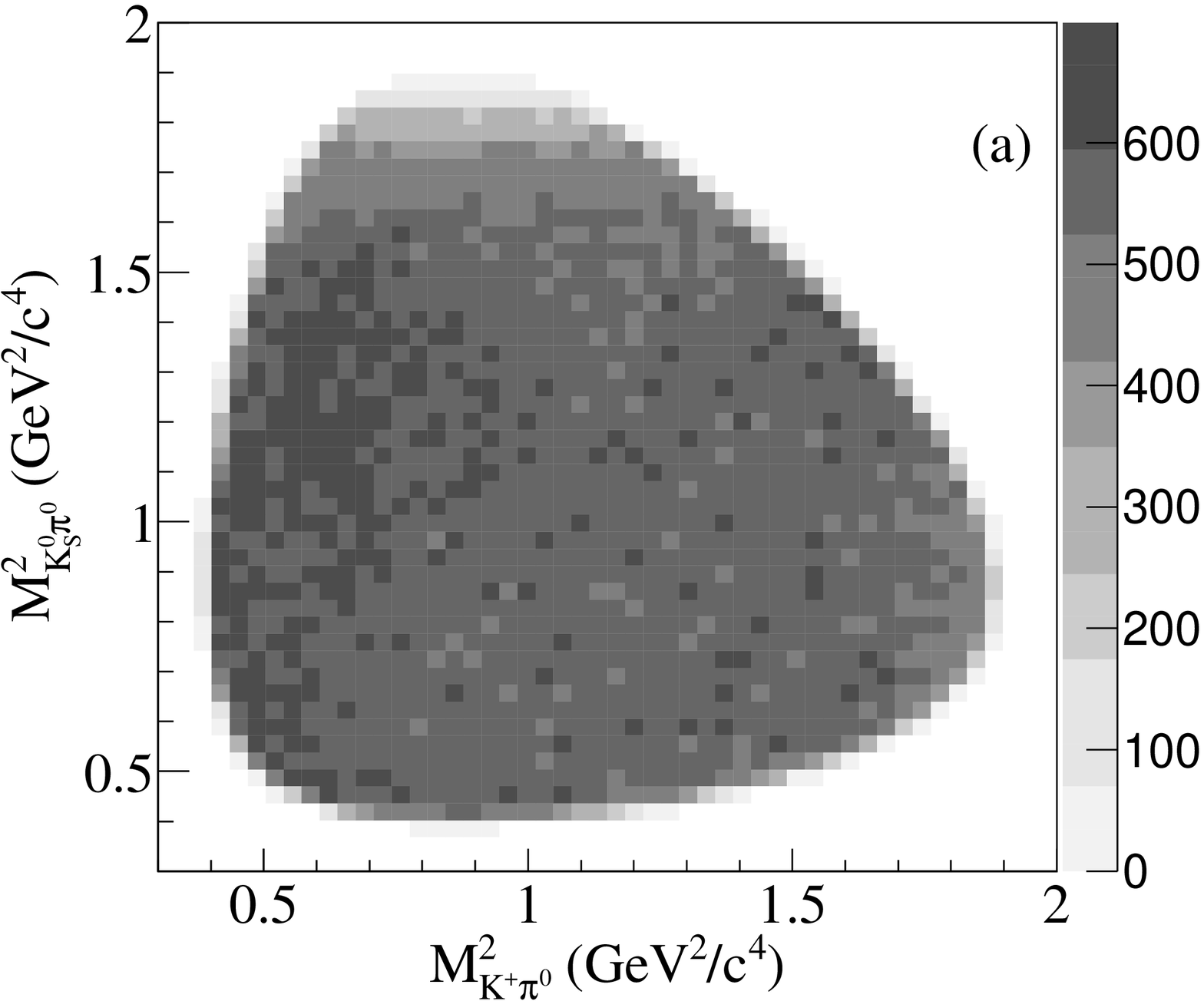}~~
\includegraphics[width=0.33\textwidth]{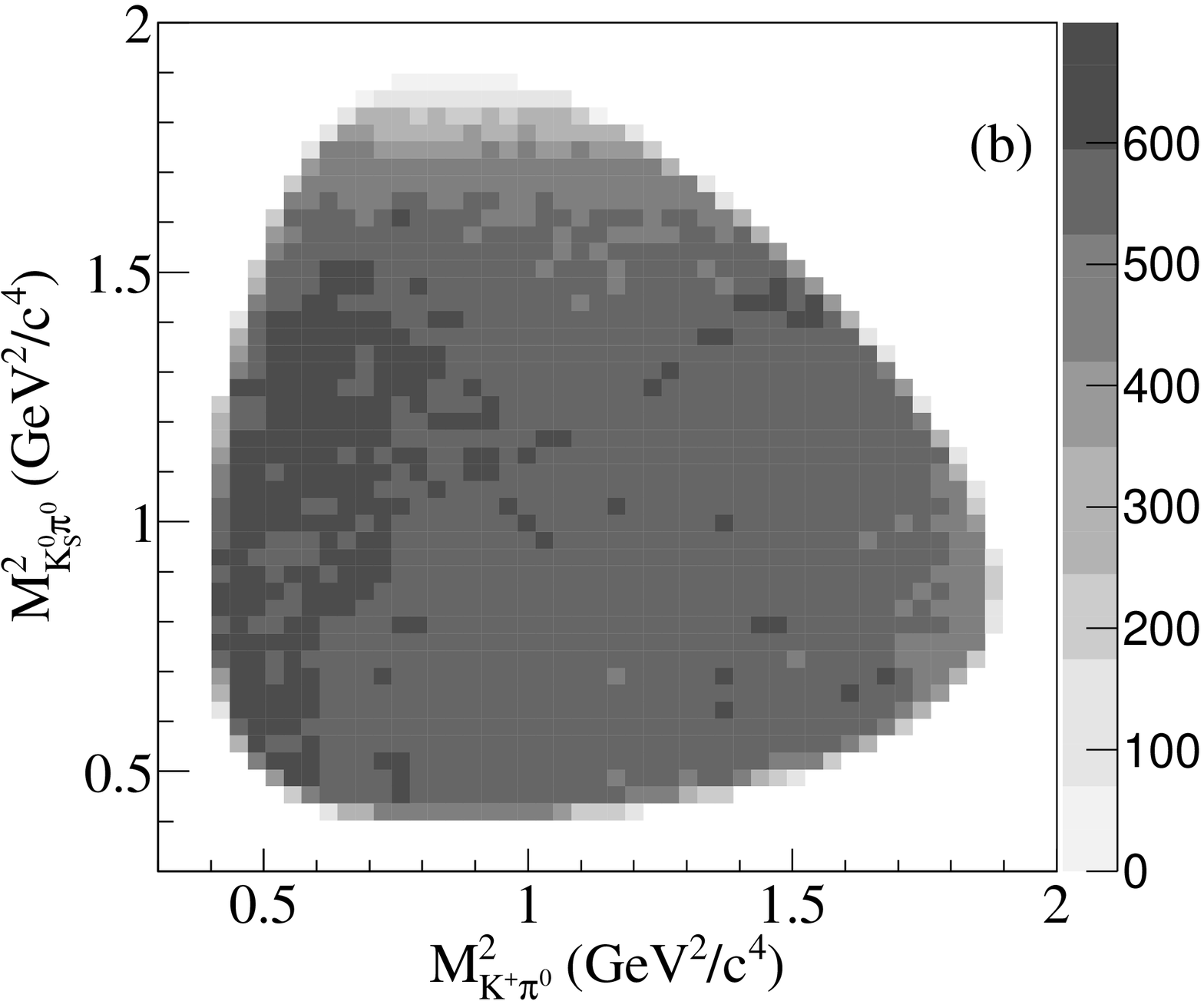}\\
\includegraphics[width=0.33\textwidth]{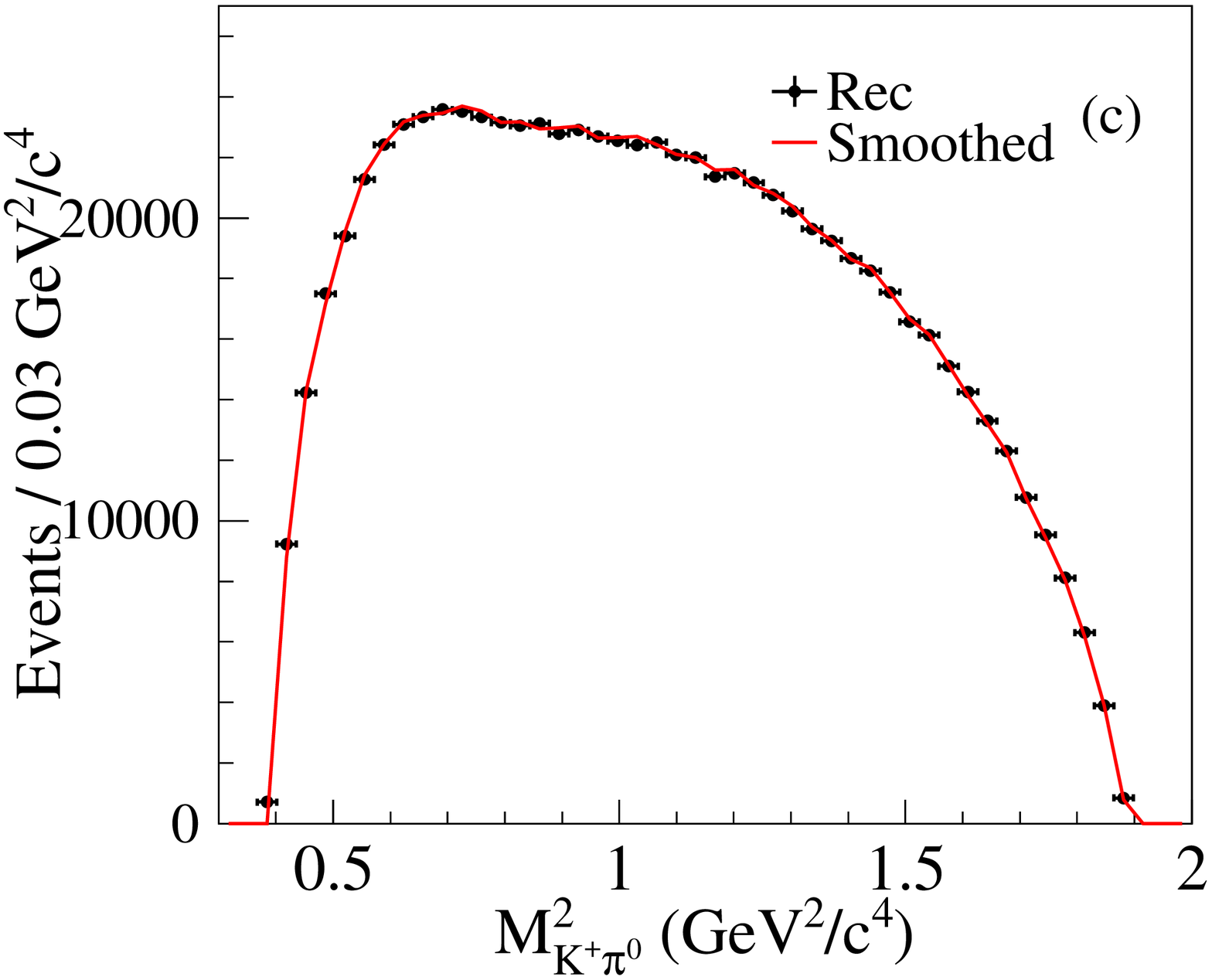}~~
\includegraphics[width=0.33\textwidth]{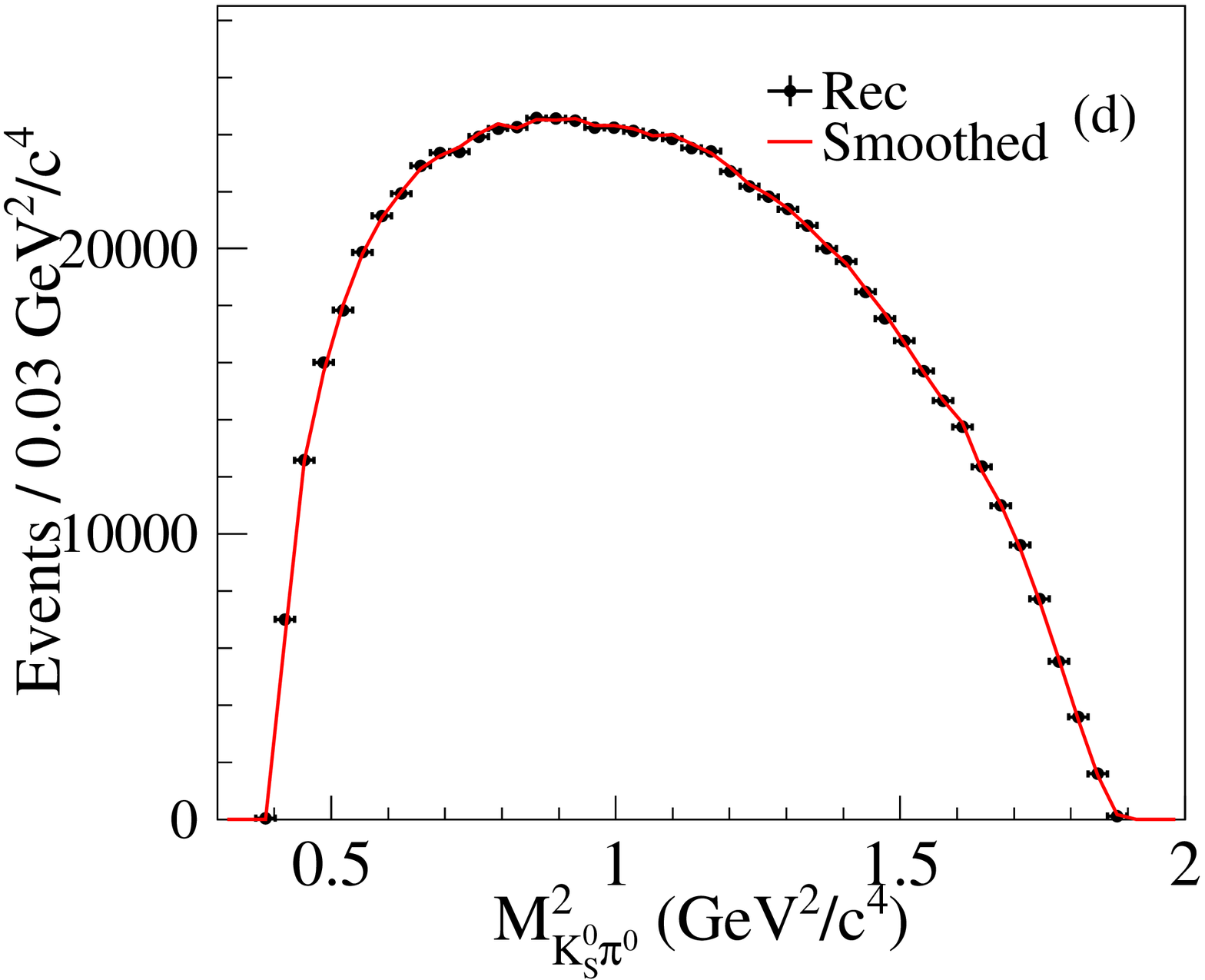}
\caption{Dalitz plot for (a) the reconstructed distribution of the PHSP MC sample and (b) the corresponding smoothed distribution. The projections on (c) $\mkppiz$ and (d) $\mkspiz$, where the dots with uncertainties denote the reconstructed and the solid lines denote the corresponding smoothed distributions.}
\label{fig:eff}
\end{figure*}

\subsection{Fit fraction and goodness-of-fit test}
The fit fraction (FF) $f_i$ for the $i$-th component is calculated with
\begin{eqnarray}
f_i=\frac{|c_i|^2\int|\mathcal{A}_i|^2\mathrm{d}x\mathrm{d}y}{\sum_{j,k}{c_jc^*_k\int\mathcal{A}_j}\mathcal{A}^*_k\mathrm{d}x\mathrm{d}y},
\label{eq:FF}
\end{eqnarray}
where the integral is calculated using the same numerical method as for the integral in Eq.~\eqref{eq:norm}. Note that the sum of
the FFs is not necessarily equal to unity due to the interferences between different components. To obtain the corresponding statistical uncertainties, 
the values of the fitted coefficients $c_i$ are randomly modified for 1000 times according to the information of the covariance matrix and the root-mean-square values of the distributions of the modified FFs are taken as the statistical uncertainties.

To examine the quality of the nominal fit, goodness-of-fit tests on three different projections of the Dalitz plot are performed using the fitted results. When calculating $\chi^2$ for each projection, an adaptive binning is adopted to ensure that the minimum number of events is larger than 10 to obey the Gaussian assumption. The $\chi^2$ value is calculated by using the number of events in data $n^k_\mathrm{data}$ and the expected number $n^k_\mathrm{fitted}$ from the nominal fit in the $k$-th bin and is formulated as
\begin{eqnarray}
\chi^2=\sum_k{\left(\frac{n^k_\mathrm{data}-n^k_\mathrm{fitted}}{\sqrt{n^k_\mathrm{data}}}\right)^2}.
\label{eq:chi2}
\end{eqnarray}

\section{Fit results}
To perform the amplitude analysis, the open-source framework GooFit~\cite{goofit} is used accelerating the fit speed using parallel processing computing. 
In the fit, the resonance $\kstarp$ is chosen as the reference, whose phase and magnitude are fixed to be one and zero, respectively.
First, the fit of data is performed with the amplitudes containing $\kstarp$ and $\kstarz$, which are clearly observed in the corresponding invariant mass spectra. 
Then, two $\mathcal{S}$-wave $K\pi$ components, $\kppizswave$ and $\kspizswave$ are included.
The statistical significances of these two components, calculated by the change of the log-likelihood values $\Delta(\mathrm{NLL})$ with and without including the component and taking into account the change of the number of degrees of freedom, are both found to be larger than $5\,\sigma$.
Besides these four components, in addition $K^*(1410)$, $K^*_2(1430)$, $a_0(980)$, $a_0(1450)$, $\rho(1450)$, $\rho(1700)$ and $(K\pi)_{\mathcal{P}\mathrm{-wave}}$ components were also tested, but their statistical significances are all lower than $5\,\sigma$, thus they are not included in the nominal fit. Here, the $(K\pi)_{\mathcal{P}\mathrm{-wave}}$ denotes the non-resonant $\mathcal{P}\mathrm{-wave}$ contribution, which is modeled with the same as vector resonances like $\kstarp$ while the dynamical function is set to be constant.

Finally, the nominal fit includes four components, $\kstarp$, $\kstarz$, $\kppizswave$ and $\kspizswave$. In the fit, the nominal masses and widths of $\kstarp$ and $\kstarz$ are fixed at the corresponding PDG~\cite{pdg2020} values. The obtained results of the magnitudes, phases $\phi$, and FFs for the different amplitudes are listed in Table~\ref{tab:nominal}, where the uncertainties are statistical only. The interference fractions between amplitudes are also listed in Table~\ref{tab:interference}.
The process $\Dp\to\kstarp\ks$ is dominant with a fraction of $(57.1\pm2.6)\%$.
The comparison of the Dalitz plots between the nominal fit and data, and the projections on $M_{K^+\pi^0}$, $M_{K_S^0\pi^0}$, and $M_{K^+K_S^0}$ are shown in Fig.~\ref{fig:nominal}.  The goodness-of-fit tests show that the $\chi^2$ values are close to one and the Dalitz-plot fit quality is good.

\begin{table*}[thp]
\begin{center}
\caption{Nominal fit results of magnitudes, phases $\phi$ and FFs for different components. The uncertainties are statistical only. The total FF is $80.9\%$. The statistical significance of each amplitude is also listed.}
\begin{tabular}{c  c c c c}
	\hline \hline
	Amplitude & Magnitude & Phase $\phi$ ($^\circ$) & FF ($\%$) & Significance \\
	\hline
	$\Dp\to\kstarp\ks$               & $1.0$ (fixed)     & $0.0$ (fixed)      & $57.1\pm2.6$ & $29.6\,\sigma$\\
	$\Dp\to\kstarz\kp$               & $0.41\pm0.04$ & $162\pm10$      & $10.2\pm1.5$ & $11.6\,\sigma$\\
	$\Dp\to\kppizswave\ks$      & $2.02\pm0.37$ & $140\pm14$       & $3.9\pm1.5$  & $5.2\,\sigma$\\
	$\Dp\to\kspizswave\kp$      & $3.14\pm0.46$ &  $-173.7\pm9.7$ &  $9.7\pm2.6$ & $7.4\,\sigma$\\
	\hline\hline
\end{tabular}
\label{tab:nominal}
\end{center}
\end{table*}
\begin{table}[thp]
\begin{center}
\caption{Interference fractions between amplitudes in units of percentage, where A denotes $\Dp\to\kstarp\ks$, B denotes $\Dp\to\kstarz\kp$, C denotes $\Dp\to\kppizswave\ks$ and D denotes $\Dp\to\kspizswave\kp$. The uncertainties are statistical only.}
\begin{tabular}{c | c c c}
	\hline \hline
	& B & C & D \\
	\hline
	A & $4.8\pm0.5$ & $0.0\pm0.0$ & $6.2\pm0.4$\\
	B &                      & $2.3\pm1.5$ & $0.0\pm0.0$\\
	C &                      &                      & $5.8\pm2.6$\\
	\hline\hline
\end{tabular}
\label{tab:interference}
\end{center}
\end{table}

\begin{figure*}[thp]\centering
\includegraphics[trim = 9mm 0mm 0mm 0mm, width=0.33\textwidth]{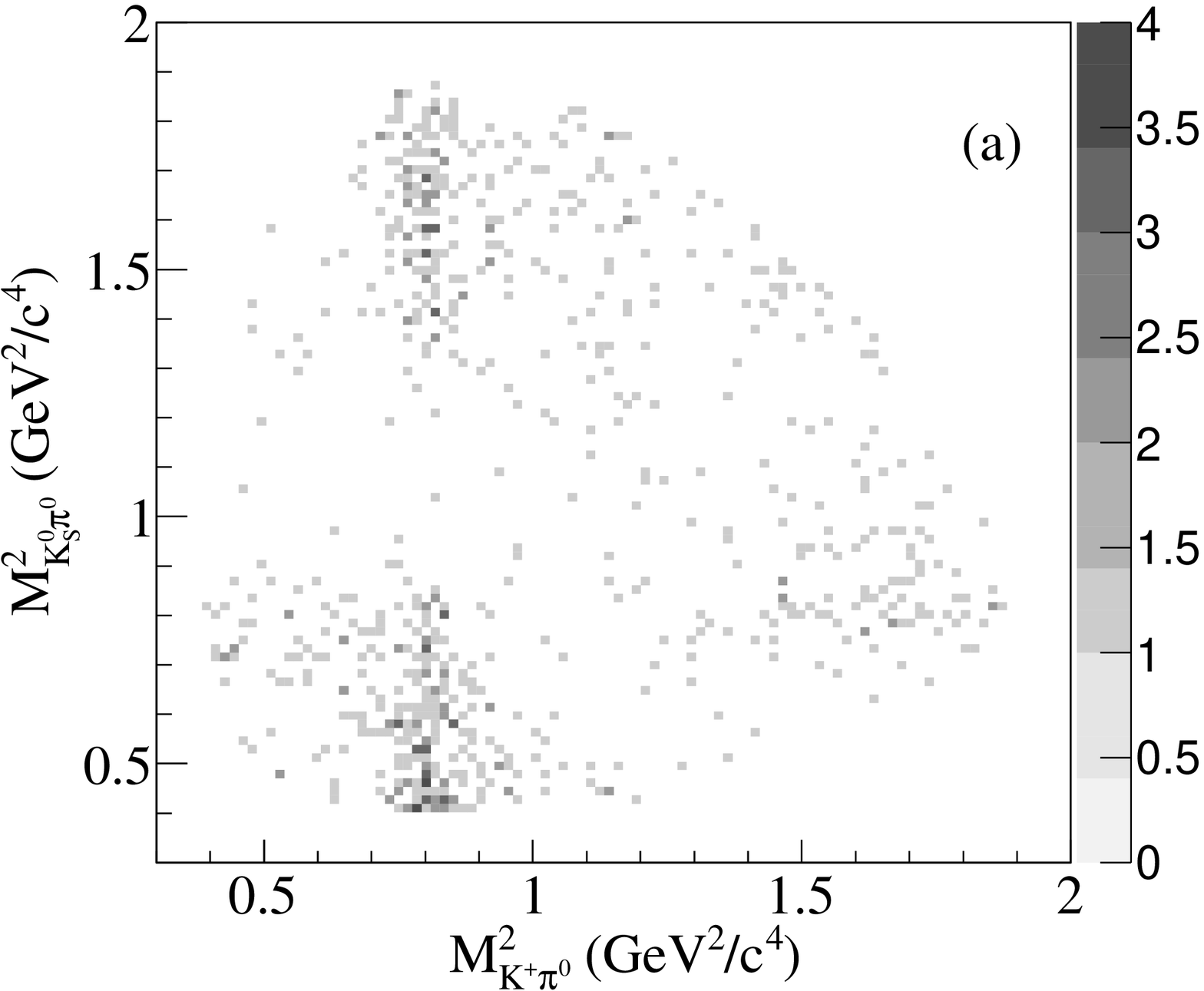}~~~
\includegraphics[trim = 9mm 0mm 0mm 0mm, width=0.33\textwidth]{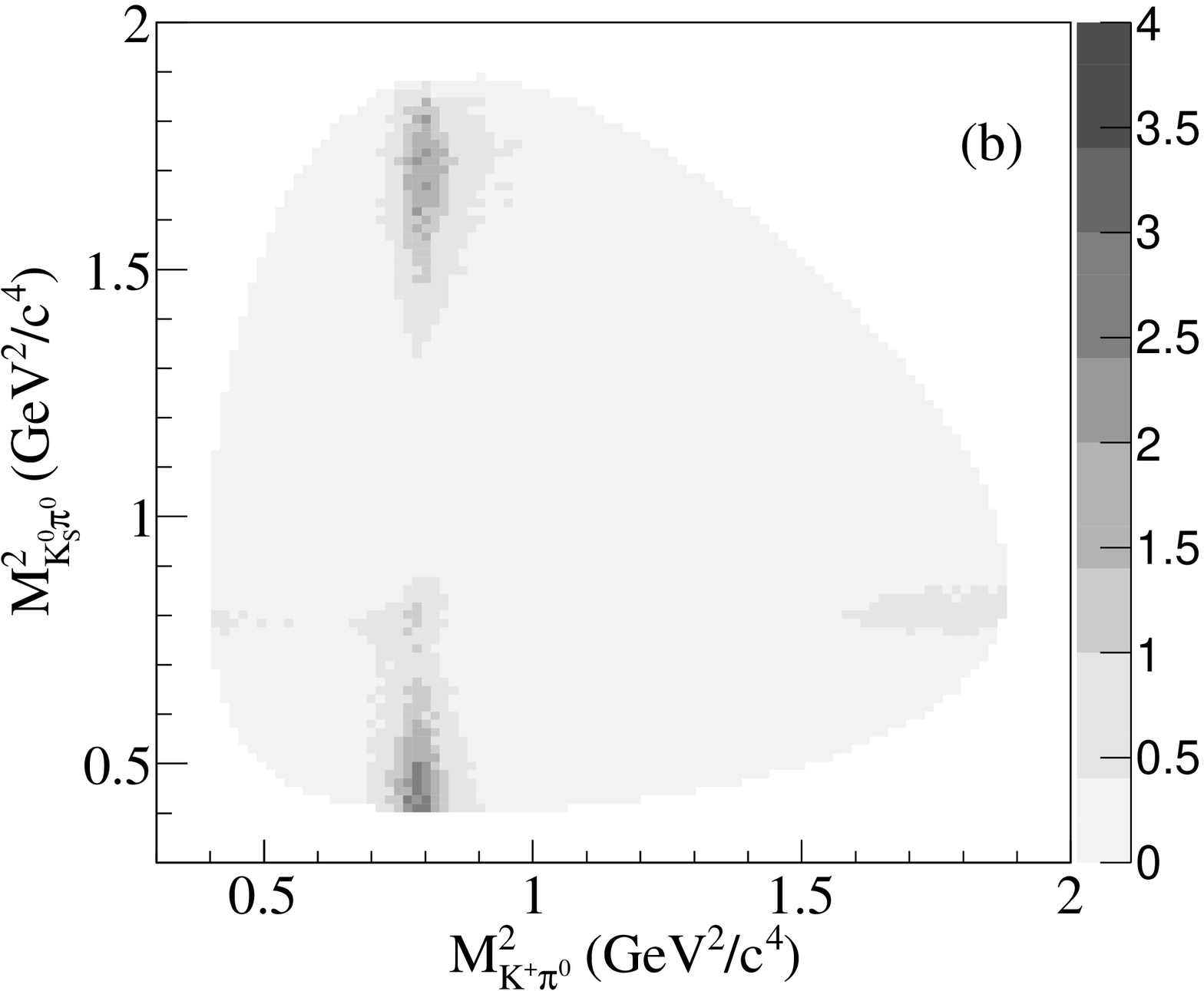}\\
\includegraphics[trim = 9mm 0mm 0mm 0mm, width=0.33\textwidth]{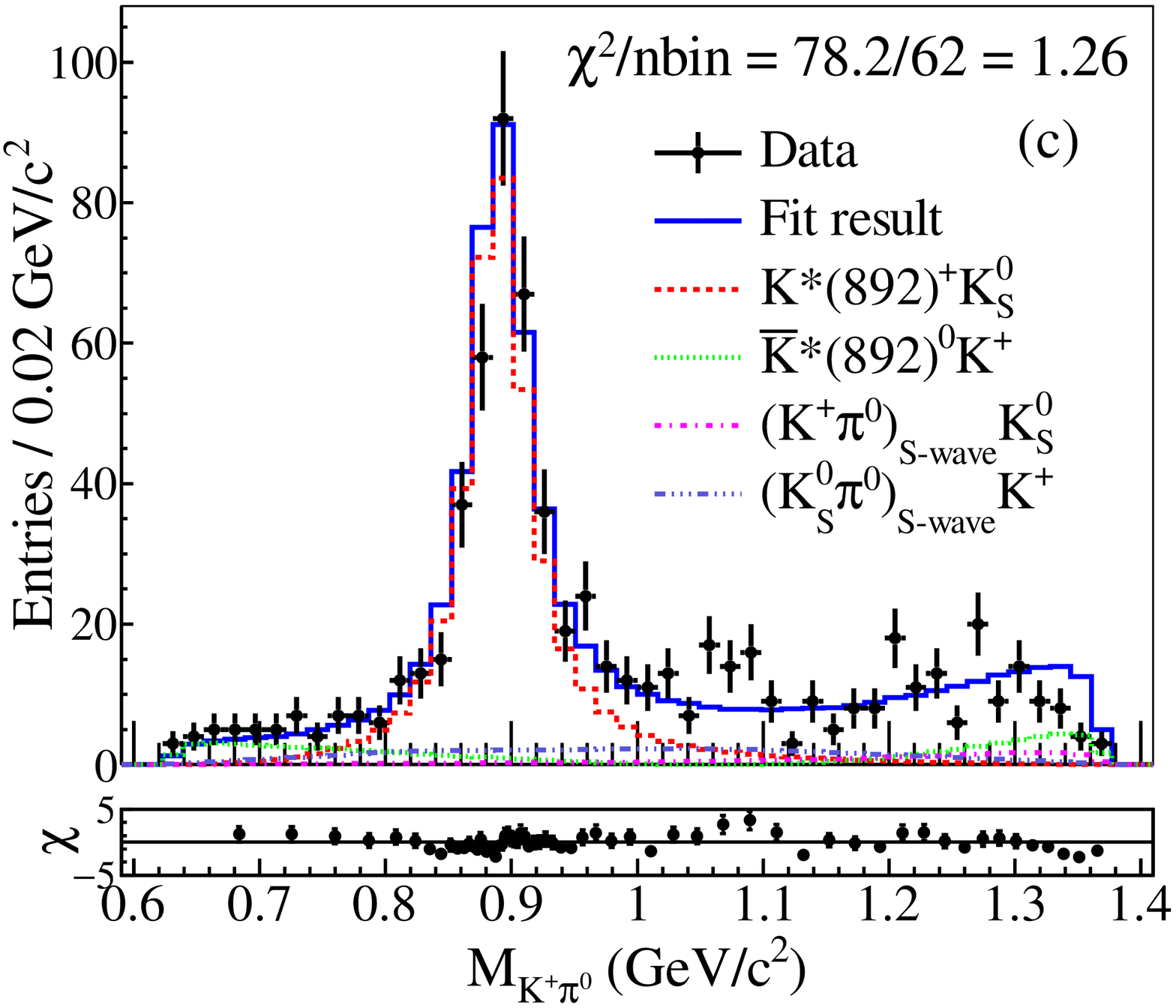}~~
\includegraphics[trim = 9mm 0mm 0mm 0mm, width=0.33\textwidth]{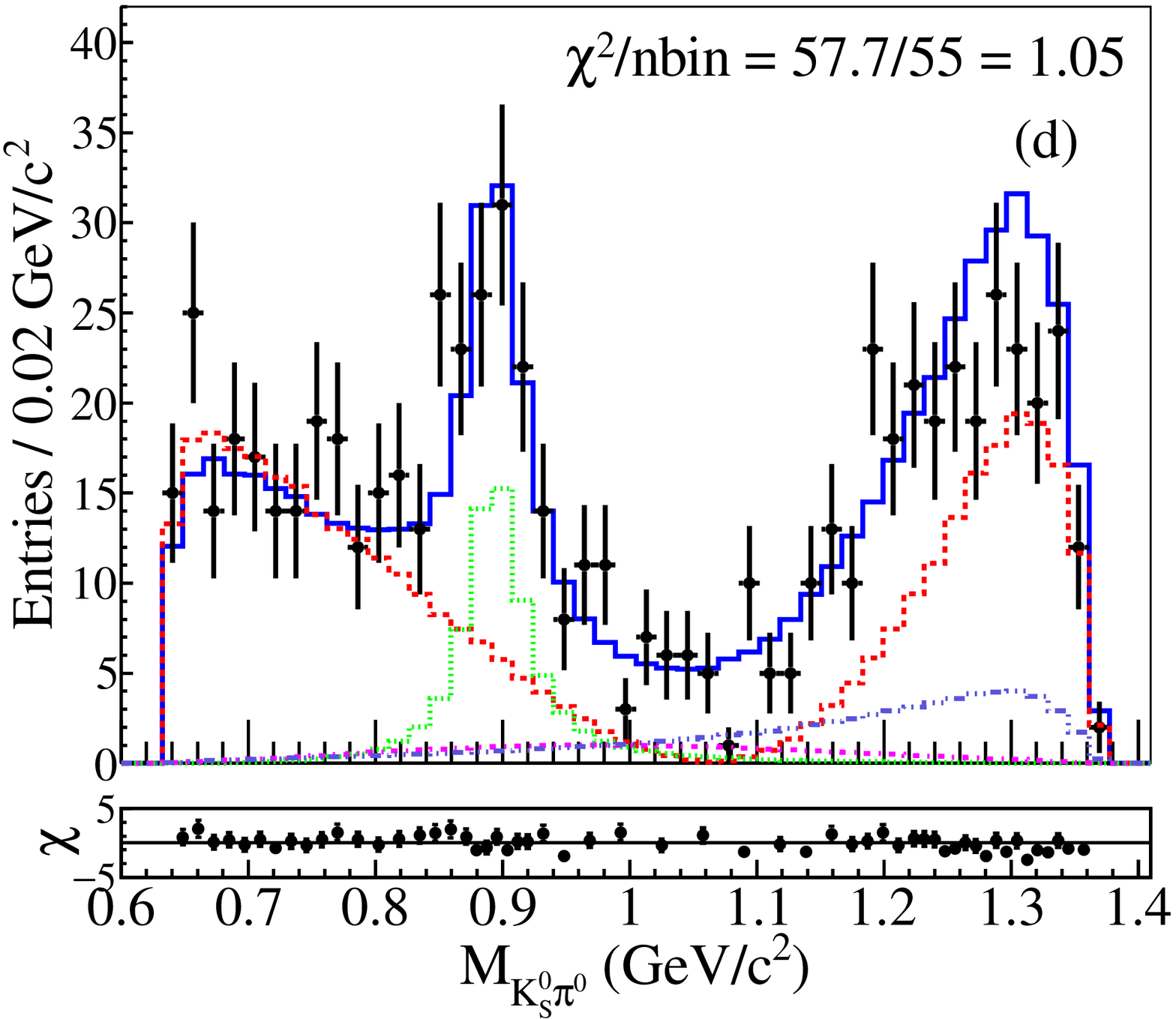}~~
\includegraphics[trim = 9mm 0mm 0mm 0mm, width=0.33\textwidth]{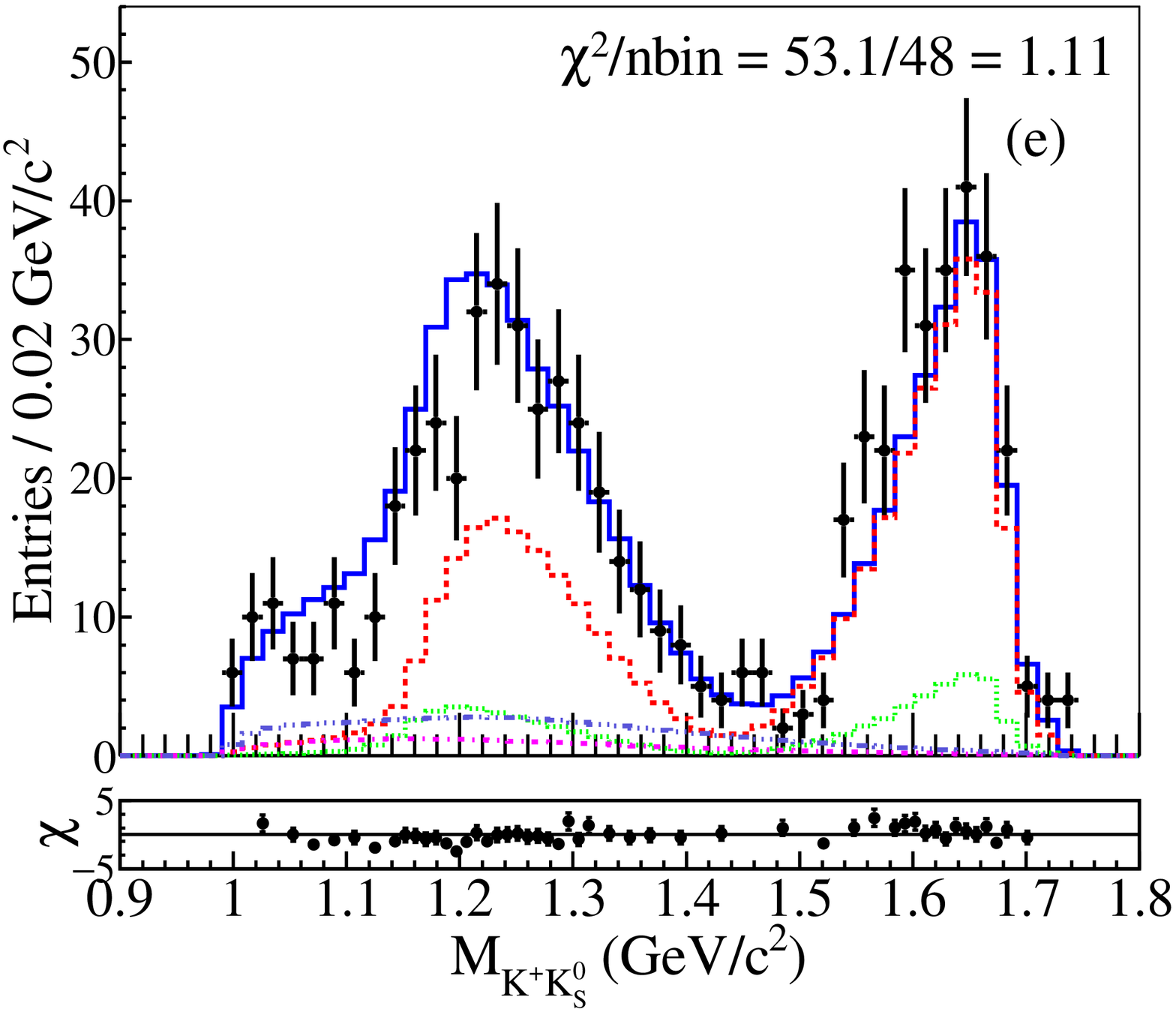}
\caption{The Dalitz-plot distributions of (a) data and (b) nominal fit, along with the projections and corresponding pull distributions on (c) $M_{K^+\pi^0}$, (d) $M_{K_S^0\pi^0}$, and (e) $M_{K^+K_S^0}$ of the nominal fit, where the black points with error bars denote data, the blue solid lines denote the fit results, and the other colored curves denote the different resonances components.}
\label{fig:nominal}
\end{figure*}

\section{Systematic uncertainty}
The systematic uncertainties for the resonance amplitudes $\Dp\to\kstarp\ks$ and $\Dp\to\kstarz\kp$ are discussed below. While the two $\mathcal{S}$-wave $K\pi$ components are included in our fit in order to improve the fit quality, in general the limited statistics of the data sample does not allow for detailed studies on these contributions so that we limit our systematic studies to the $\Dp\to\kstarp\ks$ and $\Dp\to\kstarz\kp$ results.
They are categorized into the following sources: (\romanOne) amplitude components, (\romanTwo) input parameters for resonances, (\romanThree) radius of the meson ($R_r$ and $R_D$), (\romanFour) background, (\romanFive) fit bias and (\romanSix) efficiency. The results of the systematic uncertainties for phases and FFs are summarized in Table~\ref{tab:sys_err}, where the uncertainties are given in units of the corresponding statistical uncertainties, and the total systematic uncertainties are obtained by summing up all contributions in quadrature under the assumption that different sources are uncorrelated.

\begin{table}[thp]
\begin{center}
\caption{Systematic uncertainties on the phases $\phi$ and FFs for the two resonances $\kstarp$ and $\kstarz$ in units of the corresponding statistical uncertainties. The following sources (\romanOne) amplitude components, (\romanTwo) input parameters for resonances, (\romanThree) radius of the meson, (\romanFour) background, (\romanFive) fit bias and (\romanSix) efficiency are considered. The total systematic uncertainties are obtained by summing up all contributions in quadrature.}
\begin{tabular}{c c c c c c}
	\hline \hline
	\multirow{2}{*}{Source} & {$\Dp\to\kstarp\ks$} & \multicolumn{2}{c}{$\Dp\to\kstarz\kp$}\\
	\cline{2-4}
	& FF & Phase $\phi$ & FF\\
	\hline
	\romanOne & 1.03 & 1.03 & 1.07\\
	\romanTwo & 0.08 & 0.11 & 0.12\\
	\romanThree & 1.13 & 1.07 & 1.01\\
	\romanFour & 0.01 & 0.01 & 0.01\\
	\romanFive & 0.14 & 0.05 & 0.02\\
	\romanSix & 0.38 & 0.05 & 0.07\\
	\hline
	Total  & 1.59 & 1.49 & 1.48 \\
	\hline\hline
\end{tabular}
\label{tab:sys_err}
\end{center}
\end{table}

(\romanOne) \emph{Amplitude components:}
To estimate the systematic uncertainties related to the imperfect amplitude components, several ensembles of simulated experiments (`toy MC samples') are generated based on the results of the  nominal fit with randomly added additional components from the list $K^*(1410)$, $K_2^*(1430)$, $a_0 (980)$, $a_0 (1450)$, $\rho(1450)$, $\rho(1700)$, and $(K\pi)_{\mathcal{P}\mathrm{-wave}}$.
The magnitude of the additional component is randomly distributed in the range between zero and the fitted magnitude of the $\Dp\to\kstarz\kp$ component, and its phase is randomly distributed in the range [0,\,2$\pi$).
Then, the fit procedure is repeated for each toy MC sample. From these fits, we obtain pull distributions for the fit results $\phi_{\kstarz}$, $\mathrm{FF}_{\kstarp}$, and $\mathrm{FF}_{\kstarz}$ compared to the nominal result that are well described by a Gaussian. The widths of the corresponding Gaussian functions describing the pull distributions are assigned as the associated systematic uncertainties.

(\romanTwo) \emph{Input parameters:}
In the nominal fit, the masses and widths of $\kstarp$ and $\kstarz$ are fixed to the values in the PDG~\cite{pdg2020} and the parameters of the LASS model are fixed according to Ref.~\cite{LASSpar}. To estimate the corresponding systematic uncertainties, the fit procedure is repeated by varying each of the fixed parameters by $\pm 1\sigma$. The quadratic sum of the maximum relative variations for each parameter is taken as the systematic uncertainty.

(\romanThree) \emph{Radii of the mesons:}
To estimate the relevant systematic uncertainties originating from fixing the $R_D$ value, the fits are performed with alternative $R_D$ values between $3\,\gev^{-1}$ and $7\,\gev^{-1}$.
The maximum relative variations of the fit results are taken as the relevant systematic uncertainties. 
In case of the $R_r$ value, which is one of the dominant sources of systematic uncertainty, toy MC samples are generated based on the fit results with randomly distributed $R_r$ parameters in the range$[0,\,3]\,\gev^{-1}$.
These toy MC samples are then fitted with the $R_r$ parameter fixed to the default value of $1.5\,\gev^{-1}$.
The observed pull distribution can be described by a Gaussian function. The width of the Gaussian function used to fit the pull distribution is taken as systematic uncertainty.
The quadratic sum of these two uncertainties is taken as the corresponding systematic uncertainty.

(\romanFour) \emph{Background:}
In the nominal fit, the background is neglected due to high signal purity. To estimate the associated systematic uncertainty, the NLL is alternatively constructed as
\begin{eqnarray}
\begin{footnotesize}
\begin{aligned}
-\ln\mathcal{L}=-\sum_{\mathrm{events}}\ln\Bigg[
&f\cdot\eta\left(x,y\right)\cdot
\frac{\sum_{i,j}{c_ic_j^*\mathcal{A}_i\left(x,y\right)\mathcal{A}^*_j\left(x,y\right)}}{\sum_{i,j}{c_ic_j^*I_{ij}}}\\
&+\left(1-f\right)\cdot B\left(x,y\right)
\Bigg],
\label{eq:likelihood2}
\end{aligned}
\end{footnotesize}
\end{eqnarray}
where $f$ is the signal fraction calculated from the $\mrec$ distribution in the inclusive MC sample and $B(x,y)$ is the background distribution modeled with a smoothed histogram~\cite{histpdf} constructed from the inclusive MC sample. 
After minimizing the NLL in Eq.~\eqref{eq:likelihood2} with the components of the nominal solution, the relative variation of the fit results is found to be smaller than 1\% of the corresponding statistical uncertainty. The variations are assigned as the systematic uncertainties.

(\romanFive) \emph{Fit bias:}
To understand the potential effect of the fit bias,
a series of DIY MC samples with same statistics as in data are generated. The fit procedure is repeated for each DIY MC sample and the pull distributions compared to the nominal fit result is obtained. Any deviation from a mean of zero of the pull distribution is assigned as a systematic uncertainty.

(\romanSix) \emph{Efficiency:}
Uncertainties from the efficiencies for charged particle tracking and PID,  as well as the reconstruction of $\ks$ and $\piz$ candidates have been studied with different control samples in previous works, see Refs.~\cite{sys_kaon,sys_ks,sys_piz} for examples.
To estimate the corresponding systematic uncertainties, we use correction factors comparing the estimated efficiencies in data and MC simulation, $\varepsilon_\mathrm{Data}/\varepsilon_\mathrm{MC}$, to re-weight the efficiency function. We obtain a modified efficiency $\eta'(x,y)$, which is then used instead of $\eta(x,y)$ in Eqs.~\eqref{eq:likelihood} and~\eqref{eq:norm}. 
The relative variation of the fit results using the modified efficiency is taken as the corresponding systematic uncertainty.
Additionally, the systematic uncertainties due to the $\mrec$ and $\dE$ requirements are studied by slightly shifting the boundaries within $1\,\mevcc$ and $1\,\mev$, respectively. The modified efficiency functions are obtained and the fit procedure is repeated.
Finally, the quadratic sum of the above variations is assigned as the systematic uncertainty related to the efficiency.

\begin{table*}[tbhp]
\caption{The obtained results based on the amplitude analysis. The subscript $\rm{stat.}$ and $\rm{syst.}$ denote statistical and systematic uncertainties, respectively, and $\rm{Br.}$ denote uncertainties from the quoted BF $\BR(\Dp\to\kp\ks\piz)$. For comparison, the previous experimental results~\cite{pdg2020} are also listed.  }
\begin{center}
\begin{tabular}{c|c|c}
\hline\hline
BF & This work & PDG\\
\hline
$\frac{\BR(\Dp\to\kstarp(\kp\piz)\ks)}{\BR(\Dp\to\kp\ks\piz)}$ & $(57.1\pm2.6_{\rm{stat.}}\pm4.2_{\rm{syst.}})\%$ & ---\\
$\frac{\BR(\Dp\to\kstarz(\ks\piz)\kp)}{\BR(\Dp\to\kp\ks\piz)}$ & $(10.2\pm1.5_{\rm{stat.}}\pm2.2_{\rm{syst.}})\%$ & ---\\
$\BR(\Dp\to\kstarp\ks)$ &  $(8.69\pm0.40_{\rm{stat.}}\pm0.64_{\rm{syst.}}\pm0.51_{\rm{Br.}})\times10^{-3}$ & $(17\pm8)\times10^{-3}$\\
$\BR(\Dp\to\kstarz\kp)$ & $(3.10\pm0.46_{\rm{stat.}}\pm0.68_{\rm{syst.}}\pm0.18_{\rm{Br.}})\times10^{-3}$ & $(3.74^{+0.12}_{-0.20})\times10^{-3}$\\
\hline\hline
\end{tabular}
\label{tab:result}
\end{center}
\end{table*}

\section{\boldmath Summary}
To summarize, based on an $\ee$ collision sample corresponding to an integrated luminosity of 2.93 $\ifb$ collected with the BESIII detector at $\ss=3.773\,\gev$, 
the first amplitude analysis of $\Dp\to\kp\ks\piz$ is carried out. The decay $\Dp\to\kstarp\ks$ is found to be dominant along with a small fraction of $\Dp\to\kstarz\kp$.
As listed in Table~\ref{tab:result}, the relative BFs are measured to be $\frac{\BR(\Dp\to\kstarp(\kp\piz)\ks)}{\BR(\Dp\to\kp\ks\piz)}=(57.1\pm2.6\pm4.2)\%$ and $\frac{\BR(\Dp\to\kstarz(\ks\piz)\kp)}{\BR(\Dp\to\kp\ks\piz)}=(10.2\pm1.5\pm2.2)\%$, where the first uncertainty is statistical and the second systematic.
Using $\BR(\Dp\to\kp\ks\piz)=(5.07\pm0.19\pm0.23)\times10^{-3}$, as measured by the BESIII collaboration~\cite{quote3}, $\BR(\Dp\to\kstarp\ks)=(8.69\pm0.40\pm0.64\pm0.51)\times10^{-3}$ is obtained, where the third uncertainty is due to the uncertainty on $\BR(\Dp\to\kp\ks\piz)$.
This result is consistent with previous results~\cite{quote1,pdg2020} but with a precision improved by a factor of 4.6.
It differs from the theoretical predictions in Refs.~\cite{yufusheng2014,haiyang2016,haiyang2019} by about $4\,\sigma$.
However, the result is consistent with the prediction based on the pole model~\cite{yufusheng2011}, which suffers from large theoretical uncertainty.
This indicates that the QCD-derived models need further improvements, which may lead to variations in the predicted CPV effects.
In addition, $\BR(\Dp\to\kstarz\kp)=(3.10\pm0.46\pm0.68\pm0.18)\times10^{-3}$ is obtained, which agrees well with previous measurements~\cite{pdg2020} and theoretical predictions~\cite{yufusheng2011,yufusheng2014,haiyang2016,haiyang2019}.
Future $\psi(3770)$ data samples at BESIII with larger statistics will provide more precise information about the process $\Dp\to\kstarp\ks$ and help to deepen our understanding of the internal dynamics of charmed meson decays~\cite{Ablikim:2019hff}.

\acknowledgments
The BESIII collaboration thanks the staff of BEPCII and the IHEP computing center for their strong support. This work is supported in part by National Key Research and Development Program of China under Contracts Nos. 2020YFA0406400, 2020YFA0406300; National Natural Science Foundation of China (NSFC) under Contracts Nos. 11605124, 11625523, 11635010, 11735014, 11822506, 11835012, 11935015, 11935016, 11935018, 11961141012, 12022510, 12035013, 12061131003; the Chinese Academy of Sciences (CAS) Large-Scale Scientific Facility Program; Joint Large-Scale Scientific Facility Funds of the NSFC and CAS under Contracts Nos. U1732263, U1832207, U1932101, U1932108; CAS Key Research Program of Frontier Sciences under Contract No. QYZDJ-SSW-SLH040; 100 Talents Program of CAS; Fundamental Research Funds for the Central Universities; INPAC and Shanghai Key Laboratory for Particle Physics and Cosmology; ERC under Contract No. 758462; European Union Horizon 2020 research and innovation programme under Contract No. Marie Sklodowska-Curie grant agreement No 894790; German Research Foundation DFG under Contracts Nos. 443159800, Collaborative Research Center CRC 1044, FOR 2359, GRK 214; Istituto Nazionale di Fisica Nucleare, Italy; Ministry of Development of Turkey under Contract No. DPT2006K-120470; National Science and Technology fund; Olle Engkvist Foundation under Contract No. 200-0605; STFC (United Kingdom); The Knut and Alice Wallenberg Foundation (Sweden) under Contract No. 2016.0157; The Royal Society, UK under Contracts Nos. DH140054, DH160214; The Swedish Research Council; U. S. Department of Energy under Contracts Nos. DE-FG02-05ER41374, DE-SC-0012069.


\end{document}